\documentclass[unsortedaddress,
onecolumn,showpacs,preprintnumbers,amsmath,amssymb]{revtex4}

\usepackage{graphicx}% Include Figure files     fleqn,
\usepackage{dcolumn}% Align table columns on decimal point
\usepackage{bm}% bold math
\usepackage{latexsym}
\begin{document}
\title{M-atom conductance oscillations of a metallic quantum wire}
%\author{autor1}
\author{T. Kwapi\'nski}

%\affilation{Institute of Physics, M. Curie-Sk\l odowska University, \\
%20-031 Lublin, Poland}

\address{Institute of Physics and Nanotechnology Center, M. Curie-Sk\l odowska University, \\
20-031 Lublin, Poland\\
\\
e-mail: tomasz.kwapinski@umcs.lublin.pl}

 \vspace{0.7cm}

\begin{abstract}
The electron transport through a monoatomic metallic wire connected to leads is investigated using the tight-binding Hamiltonian
and Green's function technique. Analytical formulas for the transmittance are derived and M-atom oscillations of the conductance
versus the length of the wire are found. Maxima of the transmittance function versus the energy, for the wire consisted of N
atoms, determine the (N+1) period of the conductance. The periods of conductance oscillations are discussed and the local and
average quantum wire charges are presented. The average charge of the wire is linked with the period of the conductance
oscillations and for M-atom periodicity there are possible (M-1) average occupations of the wire states.
\end{abstract}

\maketitle

\section{\label{sec1}Introduction}
The study of the electron  transport through mesoscopic systems is one of the most fundamental theoretical and experimental
problems in nanostructure physics. Interest in quantum transport in low dimensional systems has increased enormously during last
years due to its potential applications in electronics and very interesting and new phenomena we can investigate in such
structures.

The conductance of quantum wires has been investigated both experimentally and theoretically by several authors. The
experimental investigation of these structures requires tools for manipulations of atoms at the nano-scale. One way of producing
one dimensional materials of atoms is to grow them on metal surfaces \cite{jal, jal2, gam, rob}. Also using scanning tunneling
microscope (STM) technique there is possible to fabricate a few-atom chain between the tip of STM and a metallic substrate,
compare e.g. Ref. \cite{yaz}. To investigate the electron transport through the monoatomic quantum wire (QW) the mechanically
controlled break junction (MCBJ) method can be used as well \cite{smi, smi2, agr}.

The conductance of an atomic chain oscillates with a period of two atoms as the length of the wire is changed. This effect is
known as the even-odd conductance oscillation. Using the Friedel sum rule, Sim et al. \cite{sim} showed that for an odd number
of atoms in the chain the conductance was maximal while for an even number of atoms the conductance was lower. These
oscillations have been confirmed experimentally using MCBJ for Au, Pt and Ir atoms \cite{smi, smi2}. The even-odd behaviour of
the conductance can be understood in terms of the level-splitting in quantum physics and the particle-hole symmetry in the
energy spectrum when all the QW energy levels lie at the Fermi energy of the system \cite{kim}. This symmetry is also visible in
the transmittance function of the system or in the local density of states around the wire at the Fermi level \cite{lan, lan2}.
In this case the conductance can be expressed analytically, e.g. \cite{kim,zeng}  Also the calculations based on the density
functional theory for the conductance of the monoatomic Al wires have shown the four-atom oscillations of the conductance
\cite{thy}. This behaviour was explained by combining a resonant transport picture with the electron structure of the free
atomic wire (not connected to electrodes). Thygesen and Jacobsen \cite{thy} also found that the period of oscillations is
determined by the filling factor of the valence band of the infinite wire. In their calculations the half-filled band of the
infinite wire implies the even-odd oscillations while the filling factor of $0.25$ leads to a four-atom periodicity.

The influence of external time-dependent fields on the conductance of a wire connected to leads was studied theoretically using
evolution operator technique \cite{kwa}. Two regimes of frequencies of the external fields were distinguished and the six-atom
period of the conductance was found. This period was explained in terms of the structure of the transmittance coefficient of the
system. Using the Su-Schrieffer-Heeger type of Hamiltonian and Green's function technique Onipko \emph{et al} \cite{oni} have
found semi-analytical expression for the tunnel conductance for oligomers such as $C_2H_2, C_2H_4S$, etc. However, the final
result of this work does not describe any period of conductance oscillations.

%An array of quantum dots can be considered as an artificial atomic
%wire.
%The effect of electron correlations and the shape of contacts between the chain and leads on the conductance oscillations have
%been studied in Ref. \cite{mol} using density matrix renormalization group algorithm. Authors obtained that the period of the
%oscillations is determined by the inverse of the conduction band filling and the even-odd oscillations occur in the case of
%half-filling atomic levels. For non-interacting models, the electrons are mainly scattered in the contacts while electron
%correlations lead to the occurrence of additional scattering mechanism. In Ref. \cite{mol} the even-odd oscillations are
%explained only by means of the electron correlations and as a consequence the amplitude of these oscillations increases with the
%length of the chain (but for $U=0$ these oscillations disappear and the conductance is constant, independent on the length of
%the wire).

Mentioned above theoretical works are based on numerical calculations and do not present analytical formulas of the conductance
except the even-odd conductance oscillations. In this theoretical work we focus on the analytical solutions of the conductance
of a metallic wire between two reservoirs. The tight-binding Hamiltonian and the Green's function method are used in our
calculations. Analytical expressions for the transmittance are obtained and the general condition on the conductance
oscillations is found. Also the details of the conductance oscillations are discussed and the local and average charges of the
wire are analyzed.

The paper is arranged as follows. In Sec. II the transmittance function for the various parameters describing the wire is
obtained. In Sec. III the general results (the conductance and the local and average QW charges versus the length of the wire)
and discussion are presented. Conclusions are given in the last section.

\section{\label{sec1}Formalism and Analytically Solvable Cases}

In this section, the model and formalism are presented and analytical formulas allowing to calculate the transmittance and the
conductance of a wire are derived. The following notation is used in the paper: the capital letter N expresses the length of the
wire while the capital letter M means the period of conductance oscillations.

 The wire is modelled as a chain of atoms with its ends
connected to the left and right leads by the following
tight-binding Hamiltonian:

%\begin{equation}
%H = H_0 + H_V \label{eq1}
%\end{equation}

\begin{eqnarray}
H = \sum_{\vec k\alpha=L,R} \varepsilon_{\vec k\alpha} a^+_{\vec k\alpha} a_{\vec k\alpha} + \sum_{i=1}^N \varepsilon_i a^+_i
a_i+ \sum_{\vec kL,\vec kR} V_{\vec kL(R)} a^+_{\vec kL(R)}a_{1(N)} +\sum_{i=1}^{N-1} t_i a^+_i a_{i+1}+{\rm h.c.} \label{eq2}
\end{eqnarray}
where the operators $a_{\vec k\alpha}(a^+_{\vec k\alpha})$, $a_i(a^+_i)$ are the electron annihilation (creation) operators in
the lead $\alpha$ ($\alpha = L, R$) and at atomic site $i$ ($i=1,...,N$). The element $V_{\vec kL(R)}$ is the electron tunneling
coupling between the states in the left (right) reservoir and the first (last) atom and $t_i$ is the tunneling coupling between
the electron states of $i$-th and $(i+1)$-th atoms. The electron spin index is omitted and the electron-electron correlation is
neglected in the first step as we concentrate on finding the general condition for conductance oscillations.  For simplicity a
single orbital per site and nearest-neighbor interactions only are assumed.

The conductance, in the linear response regime and at zero
temperature is simply proportional to the total transmittance:
$G(E_F)={2e^2\over h}T(E_F)$. The transmittance depends on the
Fermi energy of the system and in our case can be find from the
formula \cite{dat}:

\begin{eqnarray}
T_N(E) &=& {\rm Tr \{\Gamma_L G^r \Gamma_R G^a\}}=\Gamma^L
\Gamma^R |G^r_{1N}(E)|^2 ,
 \label{A4}
\end{eqnarray}
where $G^r$ ($G^a$) is the matrix of retarded (advanced) Green functions. The only nonzero element of the matrix $\Gamma_{L}$ is
$(\Gamma_{L})_{11}=\Gamma^L$ and the function $\Gamma^L$  is expressed as follows: $\Gamma^{L}=2\pi \sum_{\vec kL}V_{\vec kL}
V^*_{\vec kL}\delta(\varepsilon-\varepsilon_{\vec kL})$ (similarly we can write for $\Gamma^R$). To find the transmittance or
conductance of the QW connected to the leads one needs to find retarded Green's function $G^r_{1N}$ which refers to the two ends
of the wire, Eq. (2). Using the equation of motion for the retarded Green's function \cite{dat} and the Hamiltonian one can
write the general matrix equation for $G^r_{ij}$, i.e. $A^N \cdot G^r=I$, where $I$ is the unit matrix and
\begin{eqnarray}
A^N\equiv A^N_{ij}=(E-\varepsilon_0)\delta_{i,j}-t_i(\delta_{i,j+1}+\delta_{i+1,j}) +i{\Gamma \over 2 }
(\delta_{i,1}\delta_{1,j}+\delta_{i,N}\delta_{N,j}) ,
\end{eqnarray}
where  $i,j\leq N$ and $A^N_{ij}$ is $N\times N$ tridiagonal symmetric matrix. The retarded Green function $G^r_{ij}$ is
obtained by finding the inverse of the matrix $A^N$, i.e. $G^r=(A^N)^{-1}$. Here, we consider the symmetric case
$\Gamma^L=\Gamma^R=\Gamma$ and assume the same electron energies of all atoms in the wire i.e. $\varepsilon_i=\varepsilon_0$ and
the same coupling strengths (hoppings) between the nearest-neighbor electron states in the wire, $t_i=t$. There are quite
reasonable assumptions as we concentrate on the linear conductance (very small or zero source-drain voltage) and take into
consideration the wire formed of the same atoms. In our calculations we put the Fermi energy as the reference energy point and
obtain the transmittance for $E=E_F$.
%Moreover, only the difference between $\varepsilon_0$ and $E_F$ ($E_F=0$) is important  and in all
%formulas in this paper the energy $\varepsilon_0$ should be treated as $\varepsilon_0-E_F$.

To find the transmittance, Eq. (2), one has to obtain the Green function $G^r_{1N}$ which can be expressed, using Eq. (3), in
the form:
 \begin{eqnarray}
G^r_{1N}={(-t)^{N-1} \over \det(A^N)}.
\end{eqnarray}
The determinant of the matrix $A^N$ satisfies the following  relation:
 \begin{eqnarray}
\det(A^N)=\det(A_0^N)+i{\Gamma} \det(A_0^{N-1}) - {\Gamma^2 \over 4 } \det(A_0^{N-2}),
\end{eqnarray}
 where
$A^N_0$ is the matrix $A^N$ for $\Gamma=0$ (non coupled wire). The determinant of the matrix $A^N_0$ can be obtained
analytically, e.g. \cite{Hu}. The above relation, Eq. (5), can be expressed in terms of Chebyscheff polynomials of the second
kind, $u_N(x)$, \cite{arf}:
 \begin{eqnarray}
\det(A^N)= t^N u_N(x) + i{\Gamma} t^{N-1} u_{N-1}(x) - {\Gamma^2 \over 4 } t^{N-2} u_{N-2}(x) ,
\end{eqnarray}
where $x=(E_F-\varepsilon_0)/2t $. Using Eq. (4) and Eq. (6) the transmittance can be written in the general form:
\begin{eqnarray}
T_N={\Gamma^2 t^2  \over {[t^2 u_N(x)-{\Gamma^2 \over 4 }u_{N-2}(x) ]^2} + t^2 \Gamma^2 u_{N-1}^2(x)}.
\end{eqnarray}
More transparent expression for the transmittance one can obtain for $|E_F-\varepsilon_0| < 2t$. In this case we have:
\begin{eqnarray}
T_N(E_F, \varepsilon_0, t)={{({\Gamma \over 2})^2 (4t^2-(E_F-\varepsilon_0)^2)} \over {[({\Gamma \over 2})^2+t^2]^2}- [({\Gamma
\over 2})^2 \cos(N-1)\phi +t^2 \cos(N+1)\phi]^2 }
\end{eqnarray}
where $\phi = \arccos(x)$ plays the role of the Bloch phase. For $|E_F-\varepsilon_0| > 2t$ the single particle energy
$\varepsilon_0$ lies beyond the wire band and this case is out of our interest. Here the transmittance is expressed only by
three parameters of the system: $E_F-\varepsilon_0, t$ and $\Gamma$.

To find the condition for M-atom conductance oscillations one has to solve the relation: $T_k=T_{k+M}$. Using Eq. (8) the
following condition can be obtained:
\begin{eqnarray}
\cos({\pi l \over M})={E_F-\varepsilon_0 \over 2 t} \,, \,\,\,\, 0<l<M
\end{eqnarray}
This equation indicates the relation between  $E_F-\varepsilon_0$ and $t$ which leads to M-atom conductance oscillations of
arbitrary length QW. It is worth to mention that for the period equals to $M$ one has $M-1$ possible solutions of Eq. (9).

Most theoretical works have studied the conductance of the quantum wire using \emph{ab initio} calculations, e.g.
\cite{lan,lan2,thy,sim}. The analytical solution for the conductance, obtained for any length of the wire, concerns mainly the
case of the even-odd conductance oscillations \cite{kim,zeng}. Using Eq. (9) one can find that the even-odd oscillations,
$T_n=T_{n+M}$, $M=2, n>0$, are observed only for $E_F-\varepsilon_0=0$. In that case the transmittance can be expressed as
follows ($k=0,1,...$):

\begin{eqnarray}
T_{2k+1}&=&1 \nonumber\\
T_{2k+2}&=&{{\Gamma^2 t^2}\over {[t^2+(\Gamma/2)^2]^2}}
\end{eqnarray}
For odd number of atoms in the wire the transmittance is maximal whereas for even ones the transmittance depends on $t$ and for
large enough $t$ the even-atom transmittance is minimal. For $t=\Gamma/2$ we do not observe even-odd oscillations because the
transmittance is maximum, i.e. $T_k=1$, for all $k$, see Eq. (10).

Next, lets find analytical formulas for the transmittance in the case of three-atom periodicity, $M=3$. The relations between
$E_F-\varepsilon_0$ and $t$ for three-atom conductance oscillations, obtained from Eq. (9), are: $E_F-\varepsilon_0=t$ for $l=1$
or $E_F-\varepsilon_0=-t$ for $l=2$, or equivalently $(E_F-\varepsilon_0)^2=t^2$. In that case the explicit form of the
transmittance of the wire can be written as follows:
\begin{eqnarray}
T_{3k+1}&=&{{\Gamma^2}\over { t^2+\Gamma^2
 }} \nonumber\\
T_{3k+2}&=&{1 \over {1+({{\Gamma}\over
{4t}})^2} } \\
%T_{3k+3}&=&{{\Gamma^2} \over
%{V^2_N(1-{{\Gamma^2}\over{4V^2_N}})^2+\Gamma^2}} \nonumber
T_{3k+3}&=&{{\Gamma^2 t^2} \over {(t^2+({\Gamma \over 2})^2)^2 }} \nonumber
\end{eqnarray}
For $k=0$ we obtain from the first relation of Eq. (11) that the transmittance of a single-atom wire, $T_1$, depends on the
coupling strength between QW atoms, $t$. It is so because here we have: $(E_F-\varepsilon)^2_0=t^2$ and for different value of
$t$ the position of $\varepsilon_0$ is also changed (the transmittance of a single-atom wire depends on $\varepsilon_0$).

Four-atom periodicity of the transmittance we find from Eq. (9) for $M=4$. The conditions for four-atom oscillations are:
 $E_F-\varepsilon_0=\pm \sqrt{2}t$, $(l=1,3)$ or
$\varepsilon_0=0$, $(l=2) $. In the first case we can write the following analytical expressions for the transmittance:
\begin{eqnarray}
T_{4k+1}&=&{{\Gamma^2}\over { 2t^2+\Gamma^2
 }}\nonumber\\
T_{4k+2}&=&{{4\Gamma^2t^2} \over {(2t^2+{{\Gamma^2}\over
{2}})^2+4t^2 \Gamma^2} }\nonumber\\
T_{4k+3}&=&{{1} \over {1+{{\Gamma^2}\over {8t^2}} }}\\
T_{4k+4}&=&{{4\Gamma^2t^2} \over {(2t^2+{{\Gamma^2}\over {2}})^2} }\nonumber
\end{eqnarray}
The second case ($E_F-\varepsilon_0=0$) corresponds to the even-odd oscillation which is the special case of every even-atom
periodicity.

The general condition for M-atom periodicity, Eq. (9), and analytical relations for the transmittance, Eqs. (11,12), are the
first main results of this paper. The explicit analytical equations concern the three- and four-atom periodicities of the
transmittance of the arbitrary length wire and are an extension of known so far, two-atom conductance oscillations \cite{smi,
smi2, sim, agr, lan, lan2, kim, zeng}.

In a similar way, using Eq. (9) we can find any atom periodicity in the wire, e.g. for $M=6$ we find: $E_F-\varepsilon_0=\pm
\sqrt{3} t$ $(l=1,5)$, $E_F-\varepsilon_0=\pm t$ $(l=2,4)$   or $E_F-\varepsilon_0=0$ $(l=3)$ and in these cases six-atom
periodicity we observe.

\section{\label{sec2}Results and discussion}

In this section we analyze the details of the periodicity of the conductance and  give another general condition for finding
periods of conductance oscillations in the atomic wire. Moreover, the quantum wire charge is discussed.

\subsection{M-atom oscillations}

The transmittance of the QW coupled to the leads, in general case, is expressed by the formula Eq. (8). For the special cases
one can use the explicit analytical equations for the transmittance, Eqs. (10-12) which describe two, three and four-atom
periods and are valid for arbitrary length wire. In our calculations we set $E_F=0$ and all energies are expressed in $\Gamma$
units. The conductance is in units of $2e^2/h$.

In the beginning, in Fig. 1 we show the conductance as a function of the energy $\varepsilon_0$ and the length of the wire
($N=1,...,15$). The coupling strength between atoms are $t=2$ - the upper panel and $t=4$ - the lower panel. The white color
represents the maximum value of the conductance ($G=1$) whilst the black color corresponds to $G=0$. The main conclusions,
concerning Fig. 1 are as follows:

\begin{enumerate}
\item{For N-atom wire we observe N maxima of the conductance (and the transmittance) e.g. for  three-atom wire, $N=3$, there are
three white fields. It is  known that a finite periodic system consisted of N cells exhibits a pattern of N narrow resonances in
each band, e.g. \cite{Sprung}. Moreover, it is visible that the conductance possesses nonzero values only in the regime of
$|\varepsilon_0|<2t$ - it is better shown for $t=2$ (the upper panel of Fig. 1). } \item{For larger value of $t$ ($t=4$, the
lower panel) peaks of the conductance are separated and are better visible then for $t=2$ (the upper panel). In the weak
coupling regime ($t=2$) the distances between the nearest conductance peaks are shorter and the values of the conductance
(between these peaks) are not equal to zero - we observe the gray (blue) color instead of black. For $N=1$ we observe the same
structure of the conductance, independent on $t$ as in that case we have the single-atom wire. } \item{The even-odd conductance
oscillations we observe for $\varepsilon_0=0$ and these oscillations are very well visible in Fig. 1 as alternately white and
black colors. Three-atom period of the conductance we find for $\varepsilon_0=\pm t$ and in our case for $t=4$ (Fig. 1, the
lower panel), we observe this for $\varepsilon_0=\pm 4$. The four-atom period of the conductance is visible for
$\varepsilon_0=\pm \sqrt{2}t\simeq\pm 5.65$, etc. One can easily find M-atom conductance periodicity of the wire in the
following way. For N-atom wire we obtain the maximum values of the transmittance versus the energy $\varepsilon_0$ and for these
values of $\varepsilon_0$, $M=(N+1)$ period of the conductance appears. For example, for $t=4$ and $N=2$ (5), the maximum values
of the conductance are for: $\varepsilon_0=\pm4$ ($\varepsilon_0=0,\pm4, \pm 4\sqrt{3}$) and in these cases $M=N+1=3$ (6)
periodicity we find. Of course the second case (six-atom period) is degenerated as it includes also the two- and three-atom
periods (for $\varepsilon_0=0$ and for $\varepsilon_0=\pm t$). Moreover, the maxima of the transmittance can be found from the
retarded Green's function by solving the following equation: $ReG^r_{1N}=0$. This conclusion is valid for the condition
$t>\Gamma/2$, which is satisfied in our case. For very weak couplings $t$, M-atom periodicity should be found from the general
Eq. (9). }
\end{enumerate}
To conclude, M-atom conductance oscillations appear for $\varepsilon_0=E_{max}$ (for $t>\Gamma/2$) where $E_{max}$ is the energy
where the transmittance of $N=(M-1)$-atom wire possesses the maximum value. The above conclusion is the second main result of
this paper.

\subsection{The special case: $M=3$}

As was shown in the previous section, the three-atom period of the conductance is observed for $E_F-\varepsilon_0=\pm t$ and in
this case the transmittance of  N-atom wire can be expressed by Eq. (11). In Fig. 2 we show the conductance versus the length of
the wire for: $\varepsilon_0=t=0.1, 0.3, 0.5, 0.7, 1$ and 4 (from very weak to strong coupling strength). The lines are
separated into the upper and lower parts and are plotted for better visualization. The most interesting conclusions concerning
three-atom periodicity of the conductance can be pointed as follows:
\begin{enumerate} \item{ The conductance
of every ($3k+1$) atom ($k=0,1,...$) gets smaller and smaller as the coupling $t$ increases. It results from the equation for
the transmittance, Eq. (11) for $(3k+1)$, where $t$ appears only in the denominator. In the same way we can analyze the other
cases. For ($3k+2$) atom the conductance increases with increasing $t$ and for ($3k+3$) atom the conductance varies depending on
the value of $t$.}
 \item{The maximum value of the conductance we observe for a) ($3k+1$)-atom in the case of
$t<\Gamma/2$, cf. the curve for $t=0.1$; b) ($3k+2$)-atom in the case of $t>\Gamma/2$, cf. the curve for $t=4$; c) ($3k+3$)-atom
in the case of $t=\Gamma/2$, cf. the curve for $t=0.5$. It is very interesting result because from the position of the maximum
value of the conductance one can conclude about the coupling strength between atoms, $t$.}
 \item{Using Eq. (11) we can find that for $t\in(\Gamma/2;\Gamma/{\sqrt{2}})$
 the conductance satisfies the following "lesser" relation: $G_{3k+1}<G_{3k+2}<G_{3k+3}$ and for $t=\Gamma/2$,
$(t=\Gamma/\sqrt{2})$ we have $G_{3k+1}=G_{3k+2}<G_{3k+3}$ , $(G_{3k+1}<G_{3k+2}=G_{3k+3})$ - cf. the curve for $t=0.5$
 $(t=0.7)$. However, it is impossible to get the "greater"
 relation for the conductance, $G_{3k+1}>G_{3k+2}>G_{3k+3}$. }
\end{enumerate}

In the same way, using Eq. (12), we can describe the four-atom period of the conductance. It is interesting that for every $t$
we never get  $G_{4k+1}<G_{4k+2}<G_{4k+3}<G_{4k+4}$ nor $G_{4k+1}>G_{4k+2}>G_{4k+3}>G_{4k+4}$. So the "greater" and the "lesser"
relations for the conductance cannot be satisfied in this case.

Analyzing the results presented in Figs. 1, 2 and relations (12,13) one can conclude that in the case of M-atom periodicity, the
maximum value of the conductance we observe for every $N=[M(k+1)-1]$ atom, e.g. for $M=3$ the conductance is maximum for $N=2,
5, 8,\ldots$ cf. the broken line in Fig. 2, the lower curves. It can be explained in terms of the transmittance, Eq. (12), where
for $t>>\Gamma/2$ we have $T_{3k+1}=0$, $T_{3k+2}=1$, $T_{3k+3}=0$. In general case one can obtain from Eq. (8) that the
transmittance $T_{M(k+1)-1}\equiv 1$ for the condition given by Eq. (9) and  $t>>\Gamma/2$.  This conclusion is the third main
result of the paper.

% \vspace{3cm}

\subsection{QW charge}
In this subsection we analyze the local and averaged charges localized in the QW. The charge at site $i$ we can obtain from the
following relation: $Q_i={-1\over \pi}\int^{E_F}_{-\infty} Im G^r_{ii}(E)dE$. In Fig 3. we show the average QW charge $\langle Q
\rangle$, $\langle Q \rangle={{\sum^N_{i}Q_i} \over N}$, as a function of the wire's length, $N$, for $t=4$ and for a)
$\varepsilon_0=0$ (dotted line); b) $\varepsilon_0=-t, +t$ (upper and lower broken lines); c) $\varepsilon_0=\sqrt{2}t$ (solid
line); d) $\varepsilon_0=\sqrt{3}t$ (thick line). Positions of $\varepsilon_0$, we consider here, correspond to the two, three,
four and six-atom oscillations of the conductance. Additionally, the crosses in Fig. 3 represent the local charge $Q_i$ at site
$i$ for $\varepsilon_0=-t$ i.e. the upper broken line is obtained by averaging the charges $Q_i$ for every $N$. For $N=1$ we
have only one cross; for $N=2$ there are two crosses but in that case $Q_1=Q_2$. In the case of $N=3$, $Q_1=Q_3\neq Q_2$ and
only two crosses are visible, and so on. One can see that the values of the local QW charges $Q_i$ are rather close to their
average value. Except for the two-atom oscillation the average QW charge, $\langle Q \rangle$, oscillates with the period
depending on the conductance oscillation period. The average charge of the wire tends to the constant value as the length of the
wire increases. This effect is visible for (b)-(d) cases (the broken, thin and thick lines). We conclude that the average charge
of the wire is linked with the period of the conductance oscillations. Here we analyze this effect in detail. The average charge
of the wire for the even-odd oscillations (two-atom period), $\varepsilon_0=0$, is 1/2 and does not depend on N (the dotted
line). For the three-atom period we find, for large N, the average QW charges equal to: 1/3 for $\varepsilon_0=+t$ (the lower
broken line) and 2/3 for $\varepsilon_0=-t$ (the upper broken line). For the four-atom period the average QW charges are: 1/4
for $\varepsilon_0=+\sqrt{2}t$ (the thin line); 3/4 for $\varepsilon_0=-\sqrt{2}t$ (not shown in Fig. 4) and 1/2 for
$\varepsilon_0=0$ (the special case of the two-atom period). These results are in agreement with the paper \cite{thy} where the
filling of the QW results in the periodicity of the conductance. According to this work, periods of the conductance oscillations
are determined by the inverse of the conductance band filling. However, in our case we have found that there are few
characteristic fillings of the wire which indicate the periodicity of the conductance, e.g. for the three-atom period: $\langle
Q\rangle=1/3$ or $2/3$.  It results directly from Eq. (9) where for M-atom period we have $M-1$ possible solutions of this
equation.

In general, for $M$-atom periodicity, the average charge of the wire can take ($M-1$) possible values: $1/M, 2/M, ... ,
(M-1)/M$; here we consider also the degenerate fillings e.g. 2/4, 2/6 etc. This effect means that the QW charge determines
unambiguously the period of the conductance oscillations whereas the knowledge about the period does not allow to indicate
exactly the QW charge. The general conclusion of this paragraph is the fourth main result of the paper.

\section{\label{sec3}Conclusions}
In summary, the transmittance and the conductance of the finite atomic wire have been calculated using Green's function method
and tight-binding Hamiltonian. For the symmetric case, assuming the same electron energies of all atoms in the wire and the same
coupling strengths between them one can write the main and new results of this paper as follows:
\begin{enumerate} \item{The analytical solution for the transmittance have been obtained for any length of the wire, Eqs. (7,8)
and also Eqs. (11,12) which describe the three- and four-atom periods of the conductance oscillations.} \item{The general
condition on any M-atom periodicity of the conductance have been obtained, Eq. (9), and discussed. The condition for
(N+1)-periodicity one can find also from maximum values of the transmittance vs. the energy of N-atom wire.} \item{Some
interesting results of $M>2$-atom periodicity have been discussed. The main and general concerns the maximum value of the
conductance. In the case of M-atom periodicity the maximum value of the conductance we observe for every $N=[M(k+1)-1]$ atom. }
\item{ For M-atom periodicity, the average charge of the wire can take ($M-1$) possible values: $1/M, 2/M, ... , (M-1)/M$ and in
this case there are $(M-1)$ possible fillings $\langle Q\rangle$ (degenerate and non-degenerate)  of the wire.}
\end{enumerate}
There are four main results of this paper. These results and also others included in the paper can be very useful for the future
experiment i.e. by analyzing the structure of the conductance vs. the length of the wire one can conclude about the coupling
strength between QW atoms, or from the distance of the maximal values of the conductance about the periodicity. Of course in the
experiment one needs to measure the conductance for infinitesimal bias across the wire, for the Fermi energy of the leads. If
$E_F=\varepsilon_0$ only the even-odd oscillation can be investigated, so one needs to change the energy $\varepsilon_0$ or the
coupling $t$ to measure the others oscillations, $M>2$. To prepare a such experiment one can use e.g. metallic wires on vicinal
surfaces \cite{jal,jal2}. One end of the wire can be coupled with a special electrode (prepared e.g. by epitaxy method) and the
second one could be the STM tip. The current can flow from the electrode to the tip through the atoms which are between them. By
changing the position of the STM tip we can change the amount of atoms between the electrode and the tip (the length of a wire).
In such wire one can change the position of $\varepsilon_0$ by applying the potential to the vicinal surface - the coupling $t$
remains unchanged. It is only one experimental possibility to measure the conductance oscillations of the wire and it is
believed that the results presented in this work will be confirmed experimentally in near future.

\section{\label{sec3}Acknowledgments}
This work has been supported by the KBN grant No. 1 P03B 004 28 and the Foundation for Polish Science.

\newpage
%%%%%%%%%%%%%%%%%%%%%%%%%%%%%%%%%%%%%%%%%%%%%%%%%%%%%%%%%%%%%%%%%%%%%%%%

\begin{figure}[h]
 \resizebox{0.4\columnwidth}{!}{
  \includegraphics{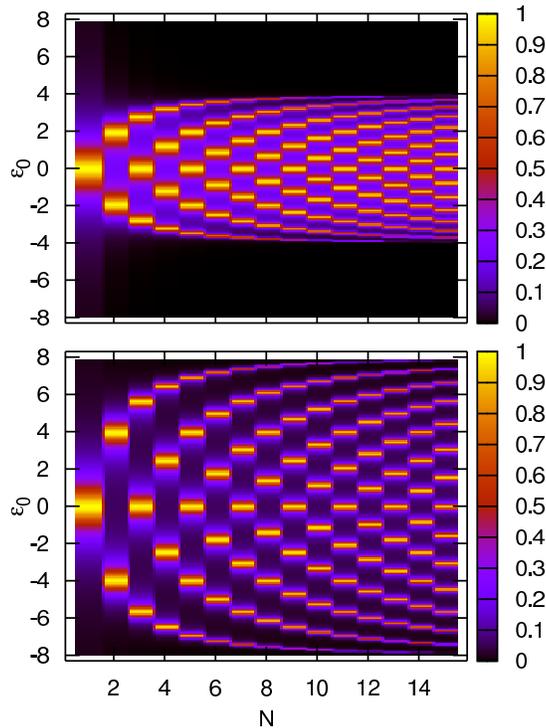}}
 \caption{\label{Fig1} (color online) The linear conductance as a function of
          the wire's length, N, and electron energy $\varepsilon_0$ for
          $t=2$ (upper panel), $t=4$ (lower panel). The black
          (light) color corresponds to $G=0$ ($G=1$); $\Gamma=1$.}
\end{figure}

\begin{figure}[h]
 \resizebox{0.4\columnwidth}{!}{
  \includegraphics{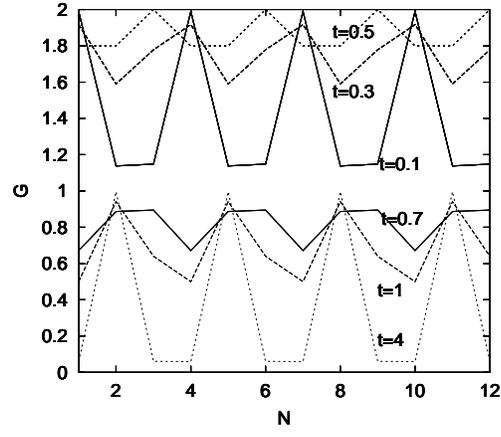}}
 \caption{\label{Fig2} The conductance versus the length of the wire N
 for $\varepsilon_0=t=0.1, 0.3, 0.5$ (the upper curves) $0.7, 1, 4$
 (the lower curves). The lines are plotted for better
 visualization. }
\end{figure}

\begin{figure}[h]
 \resizebox{0.4\columnwidth}{!}{
  \includegraphics{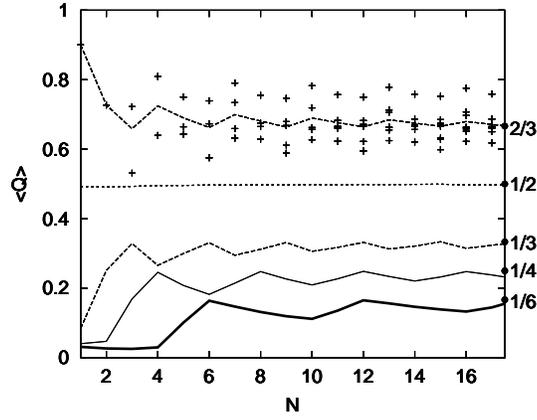}}
 \caption{\label{Fig5} The average QW charge versus the length of the wire N for
 $t=4$ and $\varepsilon_0=0$ - dotted line,
$\varepsilon_0=-4, (+4)$ - the upper (lower) broken line, $\varepsilon_0=-4\sqrt{2}$ - thin line, $\varepsilon_0=-4\sqrt{3}$ -
thick line. The crosses correspond to the local charge of $i$-atom for $\varepsilon_0=-4$. The right-side numbers represent
 fillings of the long-length wires.  }
\end{figure}
\end{document}